# Centrally Administered State-Owned Enterprises' Engagement in China's Public-Private Partnerships: A Social Network Analysis

Min Xiong
Department of Public Policy and Administration, School of International and Public Affairs
Florida International University
mxion001@fiu.edu

Travis A. Whetsell
Department of Public Policy and Administration, School of International and Public Affairs
Florida International University

Jerry (Zhirong) Zhao
Institute for Urban and Regional Infrastructure Finance, School of Public Affairs
University of Minnesota

Shaoming Cheng
Department of Public Policy and Administration, School of International and Public Affairs
Florida International University

**Abstract:** A salient characteristic of China's public-private partnerships (PPPs) is the deep involvement of state-owned enterprises (SOEs), particularly those administered by the central/national government (CSOEs). This paper integrates the approaches of resource-based view and resource-dependency theory to explain CSOEs' involvement in PPP networks. Built upon a network perspective, this paper differs from earlier studies in that it investigates the entire PPP governance network as a whole and all PPP participants' embedded network positions, rather than individual, isolated PPP transactions. Using a novel data source on PPP projects in the period of 2012-2017, social network analysis is conducted to test hypothesized network dominance of CSOEs' in forming PPPs, in light of CSOEs' superior possession of and access to strategic assets. Research findings suggest that CSOEs have a dominant influence and control power in PPP networks across sectors, over time, and throughout geographic space. It is also suggested that policy makers should reduce resource gaps between SOEs and private businesses, and only in so doing, presence and involvement of non-SOEs in China's PPPs can be enhanced.

**Keywords:** Public-private partnerships, state-owned enterprises, resource-based view, resource-dependency theory, social network analysis

**Acknowledgements:** This article has been accepted for publication in Area Development and Policy, published by Taylor & Francis.

# Introduction

The past decades have seen an increasing interest of public governments' collaborating with non-public sectors, such as private corporations and non-profit organizations, to deliver public goods and services. This is particularly the case in China in light of the sheer amount of public-private partnerships (PPPs) and the rapid growth of PPPs. By the end of 2018, there were 8,654 PPP projects, in a total amount of nearly RMB13.2 trillion, approximately USD1.89 trillion (China Public Private Partnerships Centre [CPPPC], 2018). China has become the country having the largest number of PPP projects and the highest PPP investment amount (Zhao, Su, & Li, 2018).

PPPs, in the Chinese context, refer to long-term contractual collaboration between the governments and *societal capital organizations* in various areas of public service provision. The societal capital organizations consist of state-owned enterprises (SOEs), privately-owned companies, and foreign businesses. The involvement of SOEs has become a salient characteristic of China's PPPs. SOEs are state owned or controlled, leverage ample resources, and possess extensive political and financial access, compared to their private counterparts. SOEs' strengths in controlling and influencing resource flows are amplified especially in centrally administered SOEs (CSOEs). CSOEs are affiliated with and controlled by China's the central/national government, while other SOEs are directed by provincial or municipal governments. Different from SOEs and CSOEs, private firms in China usually have limited financial capacities and access to resources, and thus higher financing costs than CSOEs (Cheng, Ke, Lin, Yang, & Cai, 2016; De Jong, Mu, Stead, Ma, & Xi, 2010; Mu, De Jong, & Koppenjan, 2011). High financing costs may render many PPP opportunities less profitable, and then restrict private firms' participation in PPPs. A total number of 981 societal capital organizations have participated in 597 national PPP demonstration projects by the end of 2017. Among these organizations there were 569 SOEs, accounting for nearly 60% of all the societal capital partners (CPPPC, 2018).

The overall question of the paper is whether CSOEs show dominating influence and control power in formation of China's PPPs because of their superior access to and control over resources. We integrate theoretical frameworks of resource-based view (RBV) and resource-dependency theory (RDT) for examining the research question. RBV serves as the foundation for understanding and analysing various participants' motivations, preferences, and priorities in PPP formation (Barney, 1991; Eisenhardt & Schoonhoven, 1996). CSOEs, other SOEs, and non-SOEs (private corporations) possess distinct assets and resources. They respectively may be sought after, through forming PPPs, by various public governments which may have unique needs for various types and/or amount of resources. In light of CSOEs' superior access to and control over strategic assets, they may then be preferred partners in PPP collaboration and partnering. RDT suggests and predicts a power asymmetry in partnering relationships if and when some participants depend heavily on other participants' resources (Pfeffer & Salancik, 2003). The overall hypothesis, informed by RVB and RDT, is that CSOEs, being resource abundant PPP partners, tend to exert a dominating influence in PPP relationship and a control power over other participants. The overall hypothesis is further decomposed and tested across industrial sectors, over time, and throughout geographic space.

Methodologically, a social network analysis (SNA) is applied for examining PPP network structures and positions in two sectors, namely 1) transportation and 2) environmental protection, in the period of 2012-2017 and across all Chinese provinces. A SNA approach is appropriate because local governments and societal capital partners are interdependent and interconnected in a governance network in which resources flow among PPP participants as they cluster to share and exchange resources. Furthermore, SNA is advantageous over existing studies in that it investigates the entire PPP governance network as a whole and examines PPP participants' embedded network positions. Existing studies generally focused on individual, isolated PPP transactions and overlook interdependence and interconnection across the PPP transactions and among the PPP participants.

The paper advances scholarly understandings of CSOEs' engagement in China's PPP formation. Theoretically, the RBV framework, which is often used to explain private sector's motivations for external partnerships (i.e., whether or not to establish a partnership), is extended to examine governments' priorities and preferences (i.e., with whom to partner). Further, the RDT approach is introduced to China's PPP context for a better understanding of asymmetrical relational connections in PPP networks as a consequence of local governments' high dependence on resources and assess possessed by CSOEs. Methodological contribution derives from the holistic network approach capturing and measuring interdependence and interconnection among PPP participants. This complements but improves existing literature which relies solely on descriptive statistics of CSOEs' involvement in individual PPP projects.

In addition, data wide, this paper uses a novel data source for exploring CSOEs' roles in PPP networks. In a response to China's national government's emphasis of developing PPPs, China's Ministry of Finance (MOF) has created a public database in 2015 for improved transparency in China's PPP development. This new database incorporates PPP projects across all the sectors, invested either by private firms or SOEs. Previous quantitative research on PPPs in China's context primarily collects data from the World Bank Private Participation in Infrastructure Project Database (Wang, Chen, Xiong, & Wu, 2018; Zhang, 2015). However, this database merely includes infrastructure projects and those mainly invested by private firms. With the novel dataset and a network analysis, the paper would also contribute to the emerging field of computational socioeconomics and policy analytics.[1] This paper also seeks to illustrate how connections between governments and societal capital organizations as well as the dominance of CSOEs are distributed over the complex geographic landscape, by integrating social network data with GIS techniques.

This paper analyses CSOEs' influence and control power in PPP networks in two different industrial sectors, over time in 2012-2017, and across all Chinese provinces. Participants' influence suggests their importance within a network measured by the extent to which they are connected with well-connected nodes. Control power refers to a participant's control over the network by bridging various participants and facilitating information exchange. The multifaceted analyses will greatly enrich our understandings of the involvement and roles of CSOEs in China's PPP formation and network.

The rest of the paper is organized as follows. The resource-based theoretical framework of PPPs is first introduced, and it is followed by an overview of China's PPP development. Hypotheses pertaining to CSOEs' dominant roles in PPP networks between sectors, over time, and across geography are developed. Based on the PPP data collected from the CPPPC, social network analysis will be carried out to measure the influence and control power of CSOEs in the PPP networks. Empirical analysis and findings on the influence and control power of CSOEs as well as the results of the robustness check are presented and discussed, and conclusions and policy options for PPP development in the future will also be developed.

## Research Background and Theoretical Framework

*PPPs through a Perspective of Resource Possession and Acquisition*

PPPs bring together resources from both public and private sectors (Hodge & Greve, 2005; Klijn & Teisman, 2003). Heterogeneous possession of, diverse acquisition for, and various degrees of dependence on resources, embodied in the resource-based view (RBV) and resource-dependency theory (RDT), characterize formation of PPP networks as well as relationships of network participants. RBV focuses on the internal analysis of organizations' strengths and weaknesses which are defined as 'resources' of a given organization (Barney, 1991). Such resources may be tangible, such as financial, physical, or human capital, but may also include intangible resources such as knowledge (Grant & Baden-Fuller, 2004). Barney (1991, p. 99) contended that the 'resources are heterogeneously distributed across firms and these differences are stable over time'. By forming strategic alliances, organizations can gain access to others' valuable resources, facilitate resource integration, and achieve mutual benefits (Das & Teng, 2000; Eisenhardt & Schoonhoven, 1996). Such strategic alliances are characterized by mutual recognition and understanding of a long-term interdependence on each partner's resources for the success (d'Alessandro, Bailey, & Giorgino, 2013; Roumboutsos & Chiara, 2010).

PPPs as strategic alliances are built upon mutual benefits to both public and societal capital partners. For governmental participants, they can acquire financial capital, technical expertise and know-how to provide public services more effectively and/or efficiently (Kivleniece & Quelin, 2012; Martin, 2018). For societal participants, in addition to projected profits, the governments usually provide political support, government sponsorships, financing assistance, government guarantees, tax exemptions or reductions, and new market opportunities to reduce their possible losses and ensure remuneration. Liu, Wang, and Wilkinson (2016), in the case study of the Beijing Metro Line 4 Project, suggested that societal capital organizations may even pursue market reputation and legitimacy at the expense of short-term profits.

Slightly from RBV which explains collaborative motives driven by resource sharing and mutual benefits, RDT emphasizes the power asymmetries based on organizational dependence on unequally distributed resources (Pfeffer & Salancik, 2003). Under the RDT, PPP participants show potential power imbalances within their strategic alliances, because their possession of resources, control over resource flows and dependence on resources differ. Klijn and Koppenjan (2016) distinguished five main types of resources: 1) financial resources such as venture capital and operating funds, 2) production resources like advanced equipment and human capital, 3) competencies such as formal authority, 4) knowledge such as technical expertise and know-how, and 5) legitimacy like political support. Any public or societal capital organization, which owns resources others do not have access to, will likely be sought after as a partner. This organization, naturally and consequently, will exert dominating influence and control power on its partners. Singh and Prakash (2010) by studying PPPs in the health services delivery in India found that the governments are influential because their partners, mainly small NGOs, are dependent on the governmental resources and have to comply with all administrative requirements and regulations.

The dominance and power imbalances among PPP participants can be indicated and measured using social network analysis (SNA), specifically, governance networks of exchanging resources within strategic alliances. Klijn and Koppenjan (2016) defined governance networks as social relations between mutually dependent actors which cluster to share and exchange resources. Based on collaborative relationships of the governments and societal partners in PPPs, a network of interdependent actors can be discerned (Hodge & Greve, 2007; Klijn & Koppenjan, 2000; Koppenjan, 2005). Each organization serve as a node and each pair-wise relationship derived from an actual contractual PPP transaction serves as an edge between two nodes. The network position of a given participant in the governance network is primarily determined by its control over and dependence on resources (Benson, 1975).

The network position of nodes exhibiting prominence, popularity, or power is normally characterized by network centrality in SNA (Wasserman & Faust, 1994). Network centrality identifies a ranking of nodes based on their connections to other nodes in the network (Lü et al., 2016). Two network centrality measures will be used, specifically, eigenvector centrality and betweenness centrality. Eigenvector centrality represents the influence of a node in the network. It assesses the extent of a node's connection particularly that to well-connected nodes (Borgatti, 2005). Hence, a node's influence is not only determined by the number of its direct neighbours but also determined by the influence of its neighbours (Lü et al., 2016). Additionally, betweenness centrality assesses the degree to which one node connect the shortest path between other nodes (Freeman, 1977). Nodes with high betweenness centrality values have high control power and act as brokers on bridges connecting other organizations in the network (Wu, Tian, & Liu, 2018).

*China's PPP Development*

The development of PPPs in China has experienced a significant fluctuation. China's legal and regulatory framework of PPPs was established in 1980s when the central government encouraged and promoted foreign investment in public infrastructure development (Zhao, Su, & Li, 2018). Promoting involvement of foreign investment at the time was to supplement local governments' own yet limited fiscal capacity in infrastructure construction in order to meet the growing demand of local economic development after China's 'reform and open-up' policy. However, such PPP engagement with foreign investment was quickly interrupted by the Asian Financial Crisis in 1998 when foreign capital was scarce (Zhang, Gao, Feng, & Sun, 2015).

The second wave of PPPs in China started in 2000, when both domestic and foreign private capital were encouraged to support rapid urbanization across Chinese cities. The Beijing Metro Line 4 project and the main stadium of the Beijing 2008 Olympic Games were two successful examples leading the second-wave of PPP projects. However, the growth of PPP projects was suspended after the Global Economic Crisis in 2008 when considerable private firms went bankrupt. Instead, local governments were encouraged to borrow through the quasi-governmental entities, i.e., local government financing vehicles, as well as to issue governmental bonds to sustain infrastructure development (Mu, De Jong, & Koppenjan, 2011). Debt financing started as a supplement to China's local governments' own-source fiscal revenues but saw a tremendous growth after the 2008 Global Economic Crisis when the central government massively expanded credits to stimulate economy (Zhao, Su, & Li, 2018).

The most recent surge of PPPs in China, starting in 2014, coincided with stringent governmental measures taken to contain China's local municipalities' alarmingly high debts. Municipal debts had been the predominant and traditional means for financing local and municipal infrastructure and services before 2014 when China tightened restriction of local governments in issuing debts. When local debts reached an alarming level and threatened fiscal sustainability in 2014, China's central government restrained local governments from borrowing and capped the local debt amount (Thieriot & Dominguez, 2015). In the meantime, PPPs have naturally become an alternative financing source for profitable public projects. Figure 1 presents trends of PPP projects and investment amounts in 2012-2017. Tan and Zhao (2019) suggested that the sudden spike of PPPs in 2015 as shown in Figure 1 is corresponding to the central government's decision in 2014 for curbing the growth of local debts in light of considerable risk of local government insolvency.

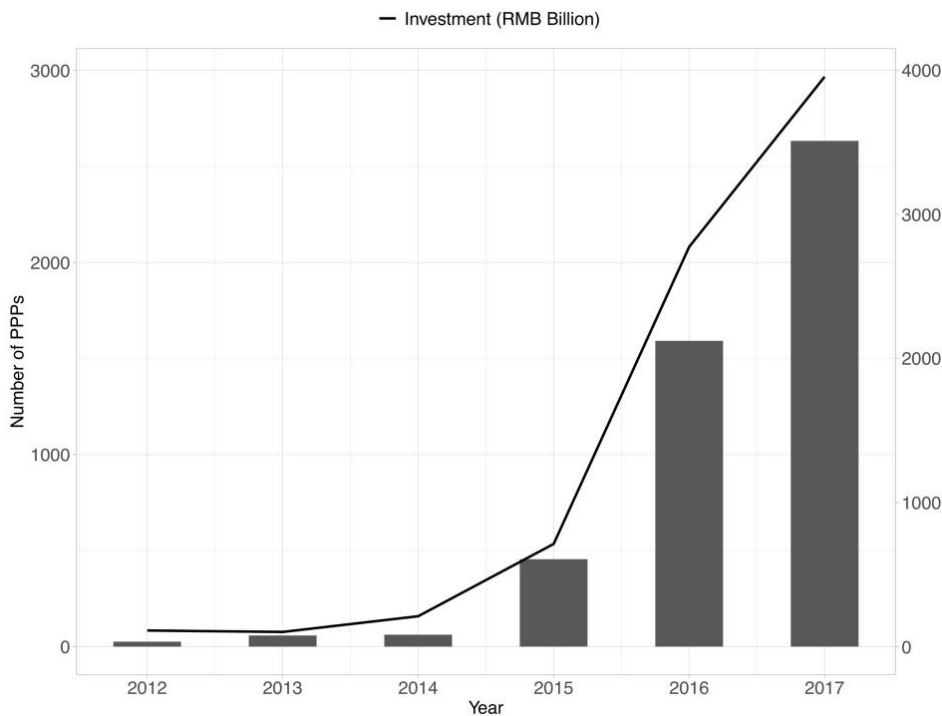

**Figure 1. Total number and investment of PPP projects across years, 2012-2017**
Data source: National PPP database by China's Ministry of Finance

*Roles of State-Owned Enterprises in PPPs*

Local governments and societal capital organizations in China exchange resources, establish strategic alliances, and achieve mutual benefits by forming PPPs. PPPs in China are characterized by the extensive involvement of SOEs. The strengths of SOEs, as major societal capital partners, are derived from the following five aspects: First, SOEs have close ties to the government. The close relationships may offer SOEs greater access to and more opportunities of PPP projects with strong cash flow, and hence may benefit indirectly local governments when they collaborate with SOEs through PPP projects (Tan & Zhao, 2019; Thieriot & Dominguez, 2015). Second, SOEs tend to have stronger financial capacities than private firms. The commercial banks prefer to provide loans to SOEs as they are endorsed by the governments (De Jong et al., 2010). Third, SOEs may be preferred partners in forming PPPs because of their experiences, expertise, human capital assets, and management skills accumulated (Mu, De Jong,

& Koppenjan, 2011). Fourth, SOEs are more stable partners in contrast to private firms. Regardless of the government transition, leadership mobility, and policy change, SOEs may be far more likely than non-SOEs to carry out the PPP projects which usually last more than ten years (Mu, De Jong, & Koppenjan, 2011; Tan & Zhao, 2019). Last, SOEs may likely enter a PPP contract with a lower profit expectation than private firms, because SOEs also bear the political and social responsibility in addition to the economic missions (Li & Zhang, 2010).

The strengths of SOEs to control and influence resource flows are amplified especially in CSOEs, those are administered and managed by the central government. By the end of 2017, there are a total of 96 CSOEs in the nation, all of which are 'extremely large firms concentrated in resource-intensive industries' (Eaton & Kostka, 2017, p. 2), such as 'finance, power and utility, petrochemical and energy, and aircraft and telecommunications' (Wang, Mao, & Gou, 2014, p. 232). Recognized as the backbone of the economy, CSOEs have greater access to strategic resources than SOEs (Huang, Xie, Li, & Reddy, 2017). In addition to the strong financial capacities, some CSOEs have set up their own engineering design and research institutes to focus on the R&D processes to improve their expertise. In contrast, local governments only have the control on limited resources which they can provide to SOEs (Li, Cui, & Lu, 2014). Hence, CSOEs have greater access to and controlling over resources, compared to SOEs and other societal capital partners.

Acquiring financial resources and overcoming financing constraints, as emphasized by the RBV, have been a predominate driver of China's local governments' enthusiasms in forming PPPs. Figure 2 presents a conceptual framework, derived from the RBV and RDT, of analysing China's PPPs. PPPs may be initiated when local governments and/or societal capital organizations seek to access each other's unique resources. On one hand, local governments gain access from societal capital organizations to financial capital, physical resources, and knowledge or know-hows which are necessary for governments to deliver public infrastructure and services. Such public infrastructure would have not been completed or accomplished in the absence of partnering with societal capital organizations. Delivering public infrastructure such as highways and railroads and harnessing such infrastructure in local economic development are primary measures for assessing local governments' effectiveness, for gaining an edge in inter-city competition, and for career promotion of local officials (Tan & Zhao, 2019; Zhu & Jiao, 2012). On the other hand, from societal capital organizations' perspective, they will not only obtain the remuneration and profits of designing, constructing, and/or operating PPP projects, but also gain legitimacy by associating and partnering with local governments (Eisenhardt & Schoonhoven, 1996; The World Bank, 2017). The access to political and policy authorities will greatly elevate societal capital organizations' competitive positions for further development. Further, as the RDT suggests, different types of societal capital organizations' heterogenous possession of resources may lead to the power imbalances among the organizations. Organizations which possess resources and/or are non-dependent on others' recourses will exert control power over the network. The relationships between societal capital organizations in different sectors will be tested in this paper, as the resource demands vary across industrial sectors.

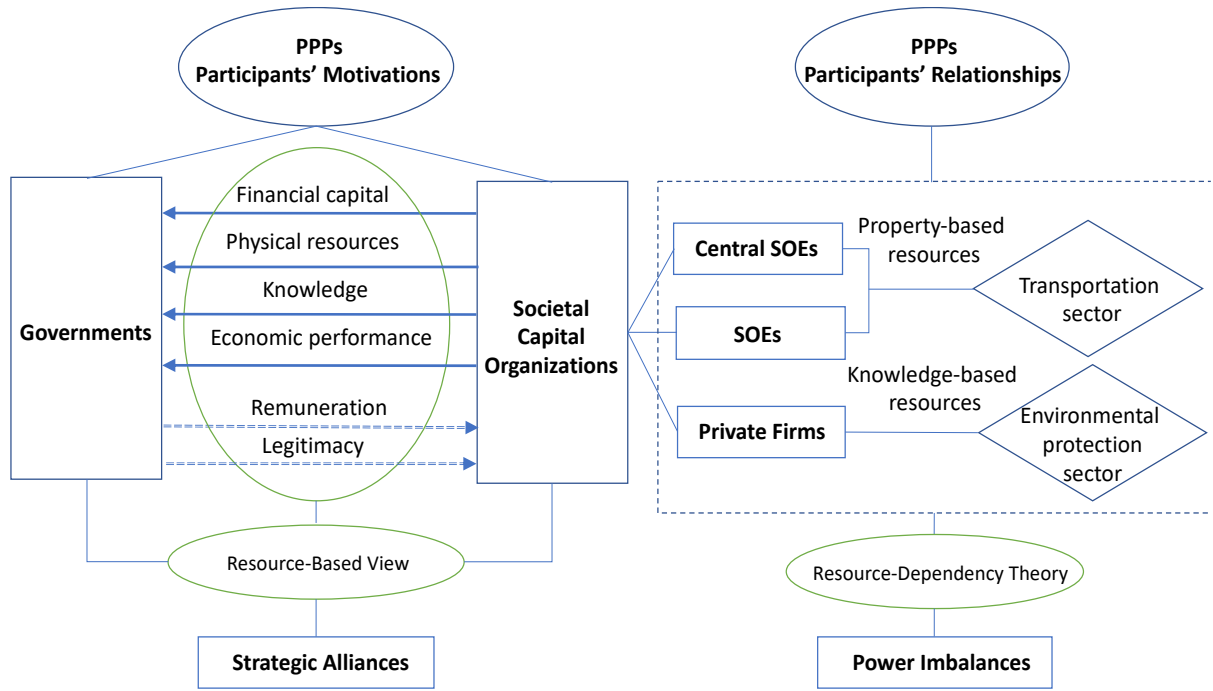

**Figure 2. Theoretical framework of PPPs in China from a perspective of resource possession**
Figure notes: Forming PPPs indicates establishing strategic alliances between local governments and societal capital organizations, derived from the resource-based view. PPPs are initiated when local governments and/or societal capital organizations seek to access each other's unique resources. Among societal capital organizations, informed by the resource-dependency theory, CSOEs and SOEs may be more influential and powerful in sectors that rely more on property-based resources (transportation), while private firms may be more influential and powerful in sectors that rely more on knowledge-based resources (environmental protection).

## Research Question and Hypotheses

The research question of this paper is whether CSOEs would exert greater influence and control power than other participants in the PPP governance network. The overall hypothesis is that CSOEs have dominant influence and control power in PPP networks.

Furthermore, the overall hypothesis is decomposed and tested in three different perspectives, i.e., across industrial sectors, throughout time, and over geography. Three sub-hypotheses are derived accordingly.

Across various industrial sectors, the demand for the types as well as magnitude of resources varies. CSOEs' influence and control power in PPP networks may vary accordingly in different industries. Das and Teng (2000) broadly classified the resources into property-based and knowledge-based. Property-based resources include financial, physical, and human capital, while knowledge-based resources refer to the expertise and skills which are usually intangible. The key distinction between property- and knowledge-based resources is the degree to which these resources can be protected from potential appropriation by alliance partners. Different types of societal capital organizations may vary in their ability to influence flows of the two distinct types of resources.

CSOEs in China are designated to produce the public goods related to the national security and economy, thus they have superior access to the financial, physical, and human capital and monopolize in the key sectors in the nation (Hubbard, 2016). CSOEs can primarily mobilize and exchange resources that are property-based and capital-intensive. Therefore, in industrial sectors that require tremendous financial, physical, and human capital, such as transportation and utilities, CSOEs would take a much stronger and influential position in networks in the property-based sectors. Private firms generally lack the property-based resources (De Jong et al., 2010; Mu, De Jong, & Koppenjan, 2011), and hence would have a stronger motivation to improve technology, optimize management, and lower costs. Private firms tend to occupy more knowledge-based resources and are likely to play a significant role in the knowledge-based sectors. Therefore, the first hypothesis pertaining to cross-sectoral variations in CSOEs' dominant roles is:

*Hypothesis 1: CSOEs' influence and control power in PPP networks are greater in sectors that rely more on property-based resources (transportation) than in sectors dependent on knowledge-based resources (environmental protection).*

Temporally, CSOEs' influence and control power in PPP governance networks may be strengthened. There may be a learning process for local governments to match their demands with resources provided by different types of societal capital organizations. By forming a strategic alliance, PPP participants aim to access the others' unique resources. The success or performance of a strategic alliance would be influenced by the learning process among partners through the experience of interactions (Zollo, Reuer, & Singh, 2002). Based on the continuous collaborations, the CSOEs' competitive advantages would become increasingly prominent. Consequently, CSOEs' dominance in PPP networks would increase over time. Therefore, the second hypothesis pertaining to variations in CSOEs' dominant roles over time is:

> *Hypothesis 2: CSOEs have greater influence and control power in PPP networks over time in both transportation and environmental protection sectors.*

From a geographical perspective, CSOEs are not evenly distributed across Chinese provinces. It is natural that CSOEs' influence and control power are greater in provinces that have higher concentration of CSOEs, mainly because of spatial proximity which reduce information search costs and encourage within-province PPP collaboration. Such geographic constraint however may be less restrictive for sectors that rely heavily on knowledge-based resources, such as environmental protection, when local governments may have greater flexibility and discretion and therefore may be able to seek non-CSOEs which have strategic access to key technology and skills. In contrast, in the sectors that rely on property-based resources like financial, physical, and/or human capitals, local government may collaborate with CSOEs and their subsidiaries based on the proximity since the property is accessible, mobile, and substitutable. Thus, the third hypothesis pertaining to geographical variations in CSOEs' dominant roles is:

> *Hypothesis 3: CSOEs' influence and control power in PPP networks are less constrained by provincial distribution of CSOEs in the environmental protection sector than in the* *transportation sector, i.e., local governments may have greater flexibility and discretion and therefore may be able to seek non-CSOEs in the environmental protection sector than in the transportation sector.*

## Data and Method

Data on PPP projects during 2012-2017 in the transportation and environmental protection sectors are collected from the public dataset of the China Public Private Partnerships Centre (CPPPC). This dataset is managed by China's Ministry of Finance (MOF), which is the official organization authorized by the central government to achieve the success of PPP projects in China.

We choose the transportation sector which is property-based and the environmental protection sector which is knowledge-based to test the above hypotheses. According to the statistics from the CPPPC (2018), the top five sectors adopting PPPs are utilities, transportation, environmental protection, urban development, and education. Transportation infrastructure projects, such as metro rails and huge bridges, are usually large scale and involve complicated technology. PPP projects pertaining to transportation infrastructure require intensive expertise, experiences, management skill, financial and human capital (De Jong et al., 2010; Mu, De Jong, & Koppenjan, 2011). The average investment amount of a transportation infrastructure PPP project is nearly RMB700 million (Shao, 2018). As a result, a large portion of infrastructure construction enterprises in China are CSOEs and SOEs which own relatively sufficient property-based resources. Taking Beijing Metro Line 4 project as an example, the societal capital partner is a joint-venture composed of Mass Transit Railway owned by Hong Kong Government, Beijing Capital Group Company Limited and Beijing Infrastructure Investment Company Limited which are two SOEs owned by Beijing Municipality Government (Liu & Wilkinson, 2013). On the contrary, PPP projects pertaining to environmental protection usually involve high-tech activities (Lee, 2010). Private firms are likely to have more opportunities and play a significant role in the environmental protection sector because they may possess technical know-hows though they may lack credentials, experience, and financial capital for infrastructure construction (De Jong et al., 2010; Mu, De Jong, & Koppenjan, 2011).

Social network analysis (SNA) will be conducted to explore and compare the characteristics of the PPP networks of the transportation and environmental protection sectors. The unit of analysis is the pair-wise PPP transaction. All governments and societal capital organizations are nodes in the network, and linkages or edges refer to the contractual interactions between the city governments and societal capital partners. The contractual relationships among various governments and societal capital organizations are extracted from actual PPP agreements and contracts, which are maintained by MOF's CPPPC dataset. Networks of the transportation and environmental protection sectors will be described independently first and then the two separate networks will be compared side by side.

Measures of network size, interconnectedness, and community structure are used to describe the whole network characteristics. The network size is measured by the total number of nodes in each network. The average degree and network density indicate the interconnectedness of the network. The average degree means the average number of edges per node in the network (Barabási, 2016). The network density is proportion of observed to potential edges in the network. A network with a high value of average degree and network density is an indication of structural cohesion (De Nooy, Mrvar, & Batagelj, 2018). However, comparison of density measures between networks of different sizes is problematic. Average degree is robust to size differences. Also, modularity is adopted to investigate the community structure in the network.

The modularity of a network is a measure of the cohesion of clusters within the network relative to the connections between clusters (Newman, 2006). A highly modular network has clusters which are disconnected from each other, while a network with lower modularity has more connections between clusters in the network.

Influence and control power of CSOEs in PPP networks are operationalized using eigenvector centrality and betweenness centrality, respectively. Eigenvector centrality and normalized betweenness centrality were calculated using the free software Gephi (Bastian, Heymann, & Jacomy, 2009). The k-core decomposition measuring CSOEs' central position in PPP networks is also included as a robust check in this study. Furthermore, Wilcoxon Rank Sum significance tests will be used to examine the differences in CSOEs' dominance across sectors as well as over time. The Wilcoxon Rank Sum test, instead of a t test, is conducted because the distribution of network centrality variables is often non-parametric and does not follow a normal distribution.

PPP networks are visualized abstractly in Figure 3 and geographically in Figure 4 to demonstrate their spatial distribution in addition to the traditional SNA. Governments and societal capital organizations in the networks are geocoded based on their geographic locations and addresses in each city. All subsidiary corporations of a societal capital organization use their own addresses not that of their parent corporations.

## Network Analysis Results

### *Characteristics of PPP Networks across Sectors*

PPP network of transportation has a larger size and is more cohesive than that of the environmental protection sector. Table 1 shows the comparative analysis of the major network measures of these two PPP networks. Compared with the network of environmental protection sector, transportation PPP network involves nearly 300 more actors and thus has much more PPP transactions during 2012-2017. Specifically, in the transportation sector, 479 city governments and 630 societal capital organizations participate in a total number of 2,130 pair-wise PPP transactions. In the environmental protection sector, there are 338 city governments and 458 societal capital partners, which together enter into 1,250 PPP contracts in total. However, it should be noted that many governments and organizations are present and overlap in both networks. On average, there are nearly three societal capital organizations per PPP contract in both sectors. A network with a high value of average degree and network density is an indication of structural cohesion. Transportation PPP network has a higher value of the average degree, which suggests that a participant in the transportation sector has realized more PPP transactions on average than that in the environmental protection sector. The density of PPP network in the environmental protection sector is a little bit higher than that in the transportation sector. However, network density is sensitive to differences in network size, where density decreases when more nodes are added. Therefore, the transportation PPP network is more cohesive that the environmental protection network. The transportation PPP network, compared to its counterpart, has a lower modularity value, as shown in Table 1. This suggests that the transportation PPP network is less partitioned into communities of densely connected nodes and hence is more interconnected and cohesive.

Table 1. Comparative analysis of PPP networks of transportation and environmental protection, 2012-2017

|  | Transportation | Environmental protection |
|---|---|---|
| **Nodes** | 1129 | 827 |
| **Edges** | 2130 | 1250 |
| **Avg. degree** | 3.773 | 3.023 |
| **Network density** | 0.003 | 0.004 |
| **Modularity** | 0.812 | 0.915 |

Figure 3 provides the visualization of these two PPP networks. As shown in Figure 3, transportation PPP network incorporates more actors (city governments and societal capital organizations in total) and more transactions than the network of environmental protection sector. Furthermore, a large densely connected community can be seen in the transportation PPP network, while the communities in the network of environmental protection sector are more partitioned. The network visualization provides support for the results presented in Table 1 that PPP network of transportation has a larger size and is more cohesive compared to the environmental protection sector.

Traditional social network analysis (SNA) visualization shown in Figure 3 is further supplemented by GIS tools to demonstrate specific geographic locations and patterns of all actors in PPP networks. Each actor in the transportation and environmental protection PPP networks has a specific geographic location, e.g., in which city a government or a societal capital organization is situated, but traditional SNA visualization randomly assigns network actors and is unable to show their spatial distribution. Given that regional disparity has become a significant feature in China's economy, such as the regional economic complexity (Gao & Zhou, 2018) and urbanization level (Chen et al., 2014), the exploration of the spatial development of PPPs would provide additional evidence to confirm the spatial differentiation. Figure 4 shows the geographic distribution of PPP transactions in the transportation and environmental protection sector respectively across Chinese provinces in the period of 2012-2017. The linkage weight (i.e., width) corresponds to the number of PPP transactions between a local government and a societal capital partner, based on their geographic locations. PPP networks of these two sectors present similar spatial patterns. Provinces located along the coast tend to have more PPP transactions in both the transportation and environmental protection sector. Compared to the environmental protection sector, provinces located in Southwest and Northwest of China adopt more transactions in the transportation sector. Also, it suggests that municipalities in the provinces with a lower level of CSOEs endowment per capita tend to go beyond provincial boundaries to collaborate with CSOEs. In other words, the collaboration between local governments and societal capital organizations is driven by municipalities' motivation of seek access to CSOEs' unique assets through participation in PPP projects.

Figure 3(a) Transportation          Figure 3(b) Environmental protection

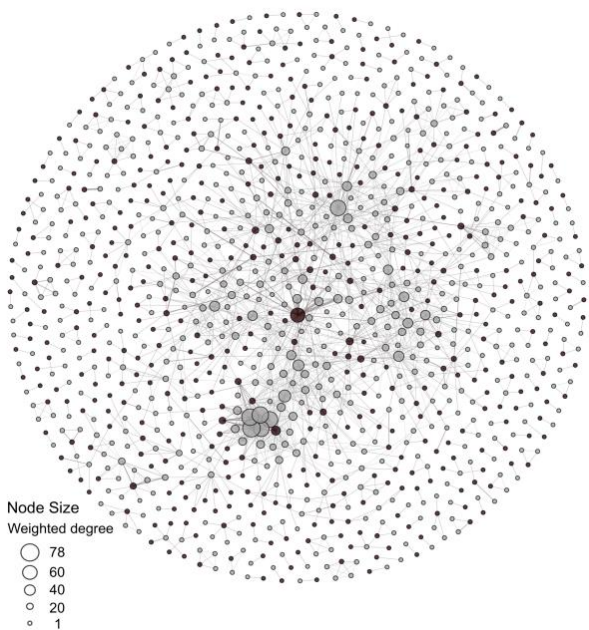
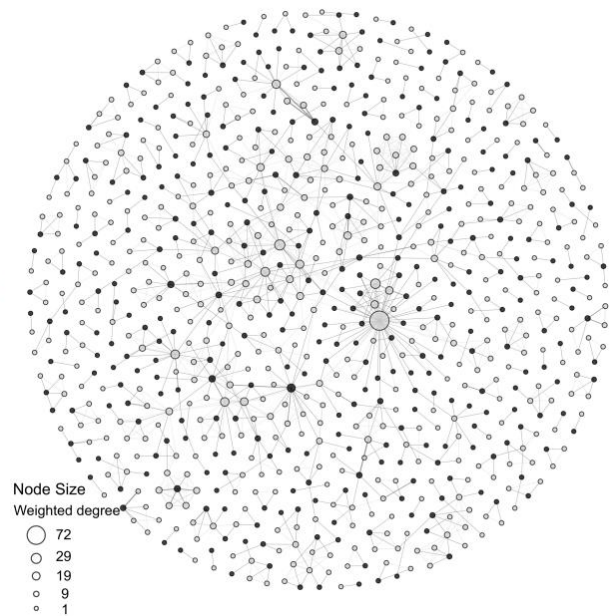

**Figure 3. PPP networks of transportation and environmental protection, 2012-2017**
Figure notes: The black nodes refer to local governments at the city level. The white nodes refer to CSOEs, SOEs, and private firms. The node size is ranked by the weighted degree (number and frequency of connections) of the node. Edges are sized by the rank of the edge weights (frequency of connections).

Figure 4(a) Transportation

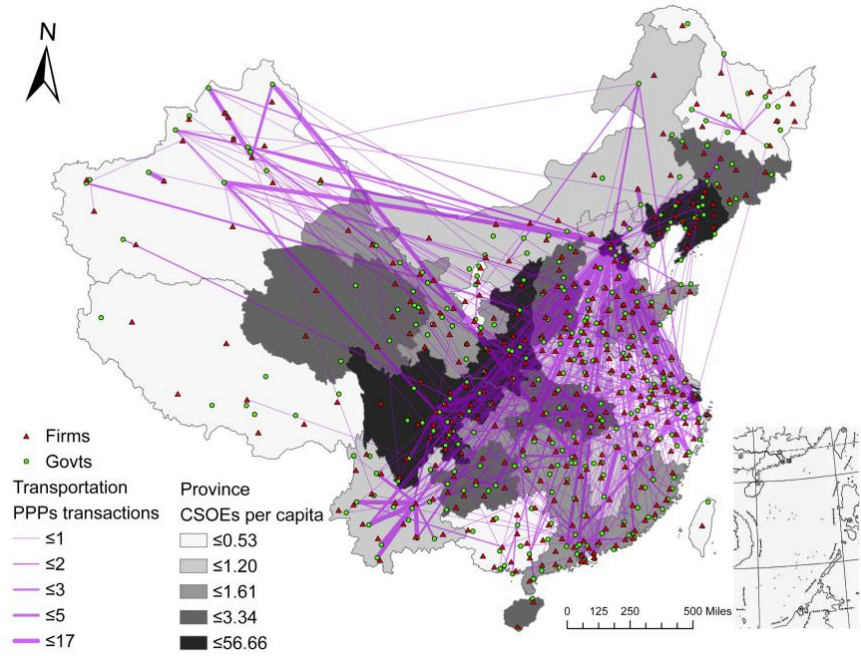

Figure 4(b) Environmental protection

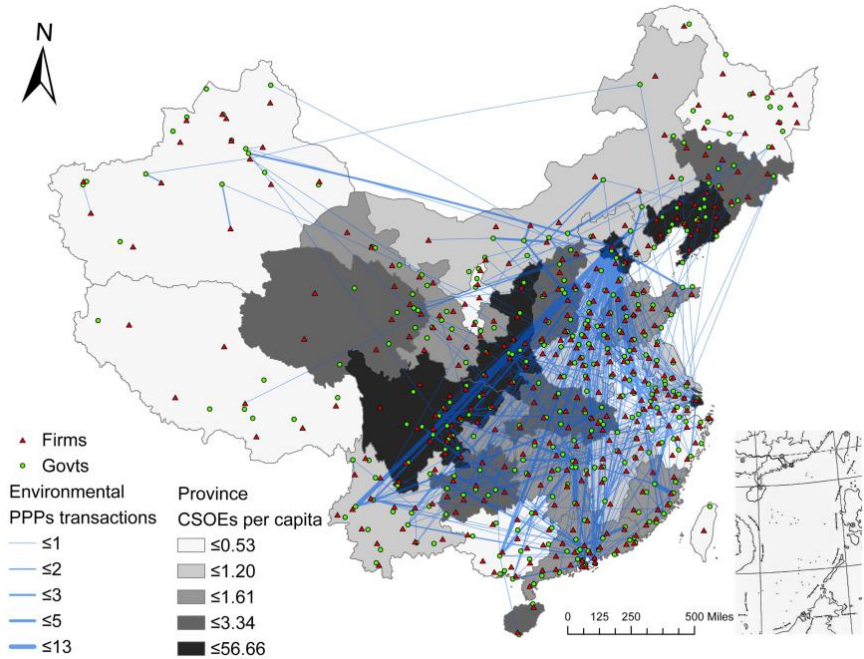

**Figure 4. Distribution of PPP transactions in the transportation and environmental protection sectors across China, 2012-2017**
Data source: National PPP database by China's Ministry of Finance

*CSOEs' Dominance in PPP Networks across Sectors*

CSOEs play a dominant role in the PPP networks of both transportation and environmental protection sectors, compared with other types of actors. Figure 5 shows the influence, which is measured by eigenvector centrality, of different types of actors in PPP networks. Eigenvector centrality assesses the degree to which an actor in the network is allied with other well-connected actors. As shown, CSOEs are the most influential actors in these two networks, compared to local government participants, other SOEs, and private firms. This suggests that CSOEs in both two sectors are dominant and exert influences over other participants. In relation to Hypothesis 1, the value of eigenvector centrality of CSOEs in the transportation sector is twice as much as that in the environmental protection sector. CSOEs therefore are more dominant and influential in the property-based transportation sector than in the knowledge-based environmental protection sector.

Similarly, as shown in Figure 6, the control power of CSOEs is much more significant compared with other types of participants, in the PPP networks of both transportation and environmental protection sectors. Betweenness centrality is adopted to measure the control power of actors in the network. Betweenness centrality denotes the number of times an actor resides on the shortest path between other actors. As indicated, the value of CSOEs' betweenness centrality in the environmental protection sector is lower than that in the transportation sector, thus suggesting a less control power over the network in the environmental protection sector.

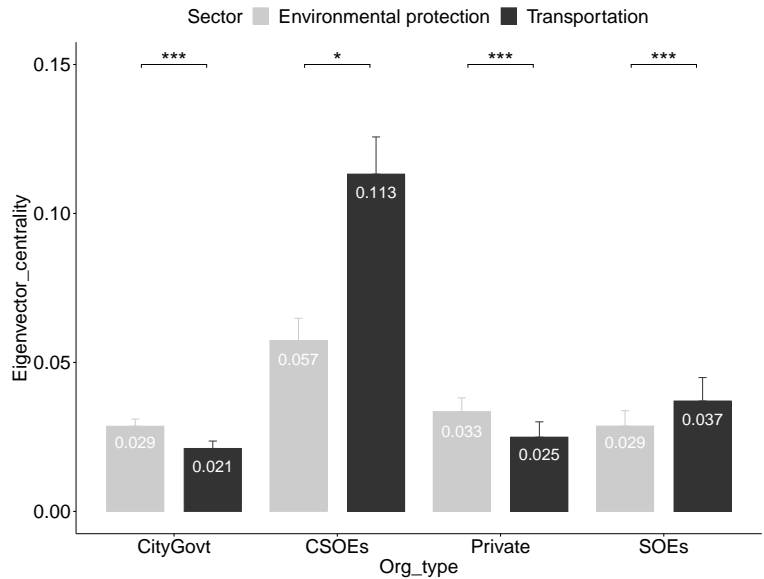

**Figure 5. Influence of different types of actors in PPP networks, 2012-2017**
Figure notes: The figure shows the comparison of eigenvector centrality scores compared with the Transportation and Environmental protection networks. Asterisks are shown on the top if the difference is significant using the Wilcoxon Rank Sum test, *p < 0.10, **p < 0.05, ***p < 0.01. Influence of actors is measured by the eigenvector centrality which assesses how well-connected an actor is to other well-connected actors in the network (Borgatti, 2005).

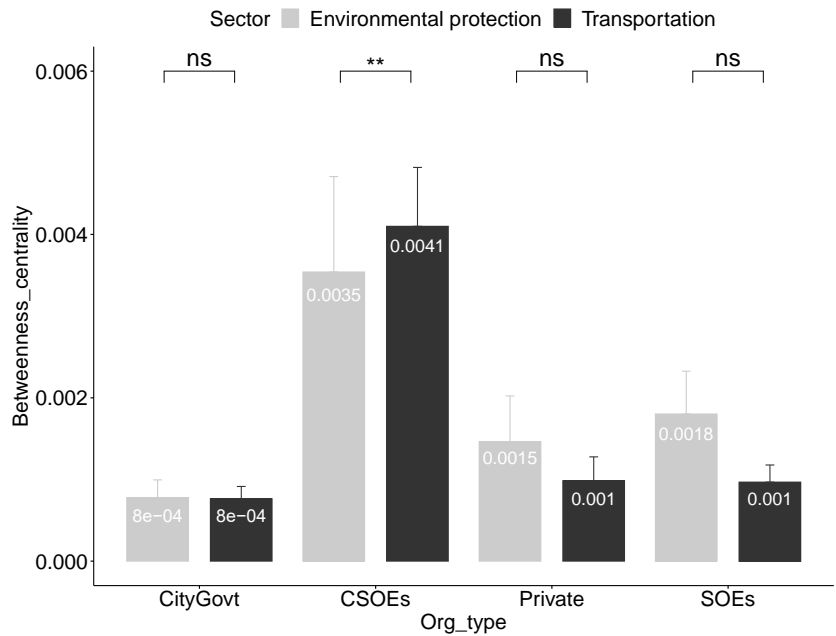

**Figure 6. Control power of different types of actors in PPP networks, 2012-2017**
Figure notes: The figure shows the comparison of betweenness centrality scores compared with the Transportation and Environmental protection networks. Asterisks are shown on the top if the difference is significant using the Wilcoxon Rank Sum test, ns= non-significant, *p < 0.10, **p < 0.05, ***p < 0.01. Control power of actors is measured by the betweenness centrality which assesses the control from an actor over the interactions between two nonadjacent actors in the network (Wasserman & Faust, 1994).

We also conduct the Wilcoxon Rank Sum test to examine whether the influence and control power of CSOEs in these two sectors are significantly different. According to the mean-comparison tests results, we find a statistically significant difference in the means of CSOEs' influence and control power in the two sectors. Therefore, the first hypothesis pertaining to cross-sectoral variations in CSOEs' influence and control power is supported.

*CSOEs' Dominance in PPP Networks over Time*

The dominant role of CSOEs varies between 2015 and 2017 in both transportation and environmental protection sectors.[2] Figure 7 shows the influence of different types of actors in transportation and environmental protection PPP networks (the mean-comparison statistics reported in Table A.1 in the Appendix). In the transportation sector, CSOEs are the most influential actor and their influence, measured by eigenvector centrality, experienced a steady increase from 2015 to 2017. However, in the environmental protection sector, CSOEs' influence decreases initially and then rises significantly in 2017. The increases of CSOEs' influence in 2017 in both sectors suggest a learning process for governments to search, screen, and match resources based on their demands and needs. Ongoing collaborations and observations of PPP transactions may strengthen CSOEs' competitive advantages and their dominant positions.

Figure 8 shows that CSOEs' control power increases in both sectors in the period of 2015-2017 (the mean-comparison statistics reported in Table A.2 in the Appendix). CSOEs have played an important brokerage role in both sectors, bridging various actors in networks for information sharing and exchange. CSOEs' betweenness centrality values in both sectors have risen steadily in 2015-2017 and higher than those of other PPP network participants, suggesting CSOEs' dominant control power in brokerage and bridging. The second hypothesis pertaining to variations in CSOEs' dominant influence and control power over time in both sectors is supported.

Figure 7(a) Transportation

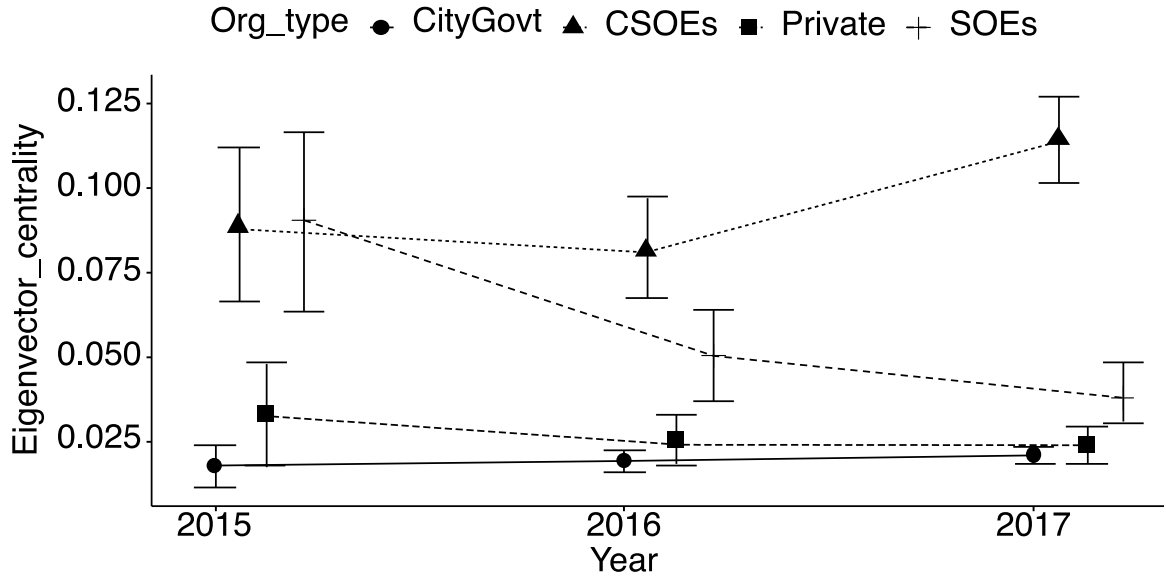

Figure 7(b) Environmental protection

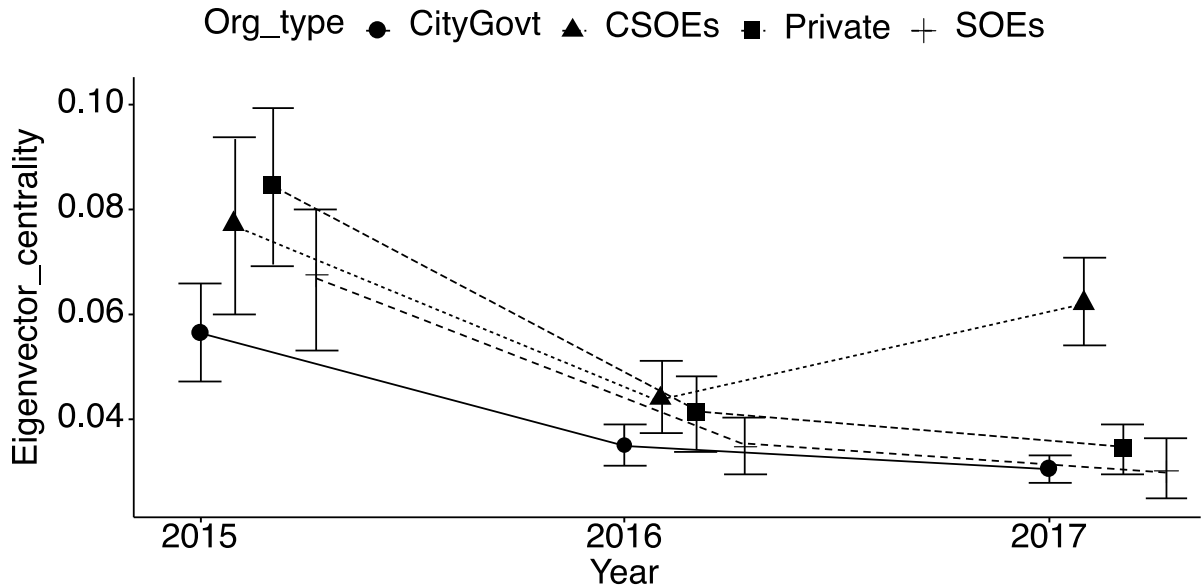

**Figure 7. Influence of different types of actors in transportation and environmental protection PPP networks, 2015-2017**
Figure notes: Influence of actors is measured by the eigenvector centrality which assesses how well-connected an actor is to other well-connected actors in the network (Borgatti, 2005).

Figure 8(a) Transportation

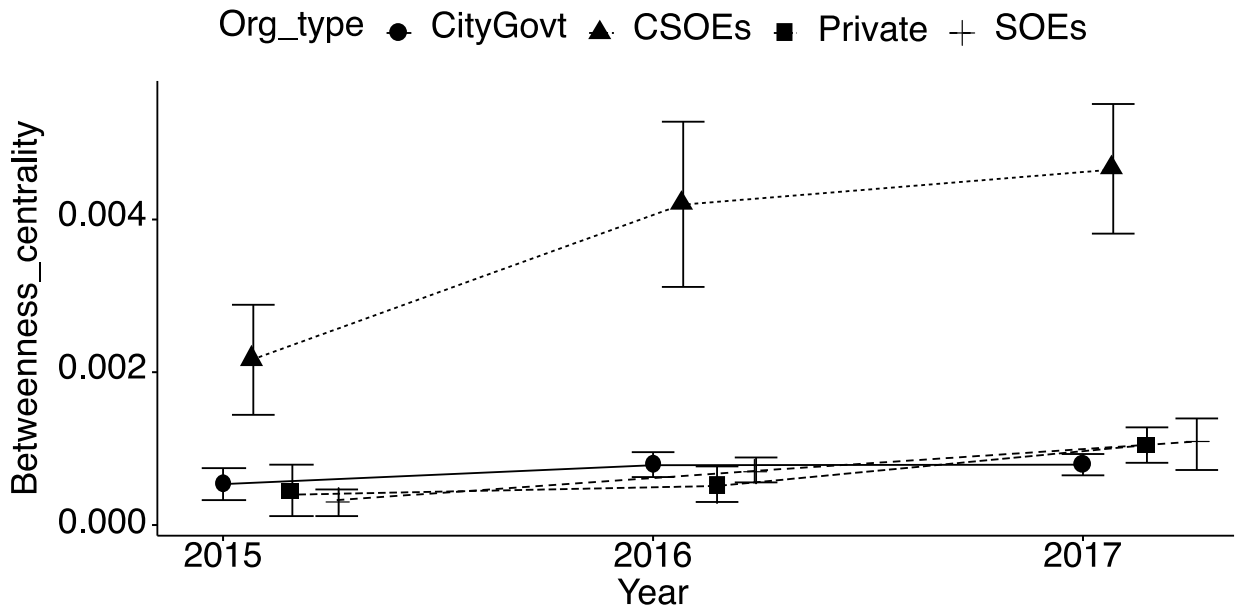

Figure 8(b) Environmental protection

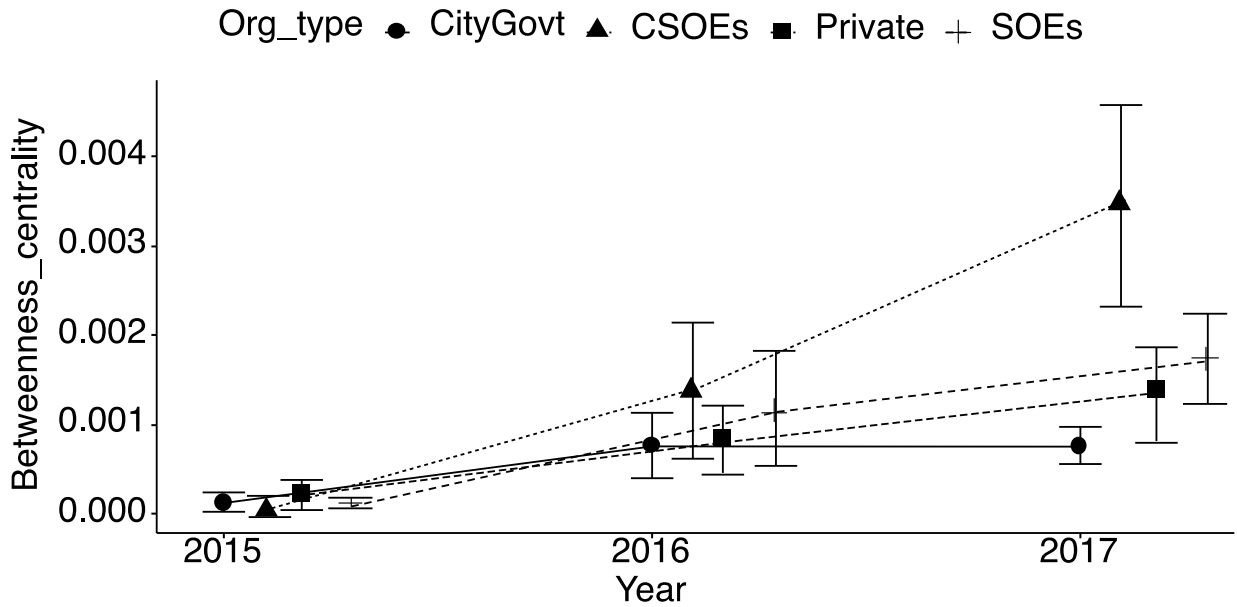

**Figure 8. Control power of different types of actors in transportation and environmental protection PPP networks, 2015-2017**
Figure notes: Control power of actors is measured by the betweenness centrality which assesses the control from an actor over the interactions between two nonadjacent actors in the network (Wasserman & Faust, 1994).

In addition, as a robustness check, we include the k-core decomposition to measure

CSOEs' dominance in PPP networks across sectors and over time. The k-core decomposition assesses the influence of a node by using the coreness as an indicator to measure whether a node is located in the core part of the network (Lü et al., 2016).Table 2 shows the results of the Wilcoxon Rank Sum test based on the coreness of different types of PPP participants in the transportation and environmental protection networks. CSOEs' coreness in the transportation sector is higher than that in the environmental protection sector. The difference between these two sectors is statistically significant. Moreover, as shown in Table 3, CSOEs' coreness becomes statistically different from other types of PPP participants in both sectors in 2017. The results for the robustness check are highly consistent with those based on the eigenvector centrality and betweenness centrality.

*CSOEs' Dominance in PPP Networks across Geography*

Spatial distribution of CSOEs' dominance is also explored. Figure 9 identifies the median centres of CSOEs' influence and control power in transportation and environmental protection sectors from 2015 to 2017. Median centre is a spatial statistic summarizing central tendency of geographically distributed values. Figure 9(a) and 9(b) are drawn based on the value of CSOEs' average eigenvector centrality and betweenness centrality by provinces in transportation and environmental protection sectors respectively. We compare them with the median centre of eight major CSOEs and all of their subsidiaries across Chinese provinces (a total number of 356 firms).[3] The median centre of major Chinese CSOEs distribution is calculated based on the number of CSOEs per capita in each province. Figure 9 shows that the median centres of CSOEs' eigenvector centrality of the transportation sector in 2015-2017 are closer to the median centre of provincial endowment of CSOEs, compared to the median centres of CSOEs' eigenvector centrality of the environmental protection sector. Similarly, the median centres of CSOEs' betweenness centrality of the transportation sector in both three years are closer to the median centre of provincial endowment of CSOEs, compared to the median centres of CSOEs' betweenness centrality of the environmental protection sector. It suggests that the spatial distribution of CSOEs' dominance in the transportation sector is more parallel to or restricted by spatial distribution of major CSOEs across China's provinces. On the contrary, PPP network participants of the environmental protection sector, though are dominated by CSOEs, are less confined by geographic distribution of CSOEs and are hence able to initiate PPP collaboration with non-CSOEs actors. Therefore, the third hypothesis pertaining to geographic variations in CSOEs' influence and control power is supported. Because of high dependence on CSOEs in the transportation sector, CSOEs influence and control power are effective to restrict transportation PPP agreement aligned with geographic distribution of CSOEs across Chinese provinces.

**Table 2. The mean-comparison test results of k-core decomposition of different types of actors in PPP networks, 2012-2017**

| Org_type | Transportation coreness | Environmental protection coreness | p | p. signif |
|---|---|---|---|---|
| City Govt | 1.979 | 1.932 | 0.306 | ns |
| CSOEs | 4.829 | 3.489 | 0.002 | *** |
| Private | 2.452 | 2.479 | 0.001 | *** |
| SOEs | 2.689 | 2.417 | 0.58 | ns |

Table notes: The table shows the comparison of k-core decomposition compared with the Transportation and Environmental protection networks, ns= non-significant, *p < 0.10, **p < 0.05, ***p < 0.01.

**Table 3. The mean-comparison test results of k-core decomposition of different types of actors in PPP networks, 2015–2017**

| Year | Group1 | Group2 | Transportation | | Environmental protection | |
|---|---|---|---|---|---|---|
| | | | p | p.signif | p | p.signif |
| 2017 | Govt | CSOEs | 8.85E-41 | **** | 7.94E-16 | **** |
| 2017 | Govt | Private | 7.33E-05 | **** | 1.19E-10 | **** |
| 2017 | Govt | SOEs | 2.84E-07 | **** | 5.62E-05 | **** |
| 2017 | CSOEs | Private | 2.69E-21 | **** | 5.26E-06 | **** |
| 2017 | CSOEs | SOEs | 4.76E-18 | **** | 2.16E-05 | **** |
| 2017 | Private | SOEs | 0.282 | ns | 0.426 | ns |
| 2016 | Govt | CSOEs | 2.18E-26 | **** | 1.50E-08 | **** |
| 2016 | Govt | Private | 0.001 | *** | 1.52E-07 | **** |
| 2016 | Govt | SOEs | 1.43E-05 | **** | 2.91E-06 | **** |
| 2016 | CSOEs | Private | 6.71E-13 | **** | 0.051 | ns |
| 2016 | CSOEs | SOEs | 1.68E-09 | **** | 0.474 | ns |
| 2016 | Private | SOEs | 0.357 | ns | 0.318 | ns |
| 2015 | Govt | CSOEs | 9.47E-13 | **** | 0.0004 | *** |
| 2015 | Govt | Private | 0.002 | ** | 0.005 | ** |
| 2015 | Govt | SOEs | 0.001 | *** | 3.16E-05 | **** |
| 2015 | CSOEs | Private | 2.42E-05 | **** | 0.151 | ns |
| 2015 | CSOEs | SOEs | 0.013 | * | 0.885 | ns |
| 2015 | Private | SOEs | 0.471 | ns | 0.147 | ns |

Table notes: The table shows the comparison of k-core decomposition of different types of actors in transportation and environmental protection PPP network, ns= non-significant, *p < 0.05, **p < 0.01, ***p < 0.001, ****p < 0.0001.

## Conclusions and Policy Implications

Informed by the resource-based view (RBV) and resource-dependency theory (RDT), this paper examines whether CSOEs would exert influence and control power over other participants in the PPP network. Based on the data on PPP projects during 2012-2017 in the transportation and environmental protection sectors, a social network analysis (SNA) is conducted to characterize PPP network structures across sectors and investigate CSOEs' dominance in PPP formation and governance. Three hypotheses are developed in line with CSOEs' influence and control power across sectors, over time, and throughout geographic space.

The SNA results show that PPP network of the transportation sector is larger in size and is more cohesive than that of the environmental protection sector. The eigenvector centrality and betweenness centrality are used to measure the influence and control power of CSOEs respectively in the PPP networks. Corresponding to the first hypothesis, the influence and control power of CSOEs are more significant in the transportation sector than in the environmental protection sector. This is because, as predicted by RBV, transportation PPPs, compared to those in the environmental protection sector, have a higher level of reliance and dependence on resources and assets possessed by CSOEs, such as credentials, financial and human capital.

Figure 9(a)

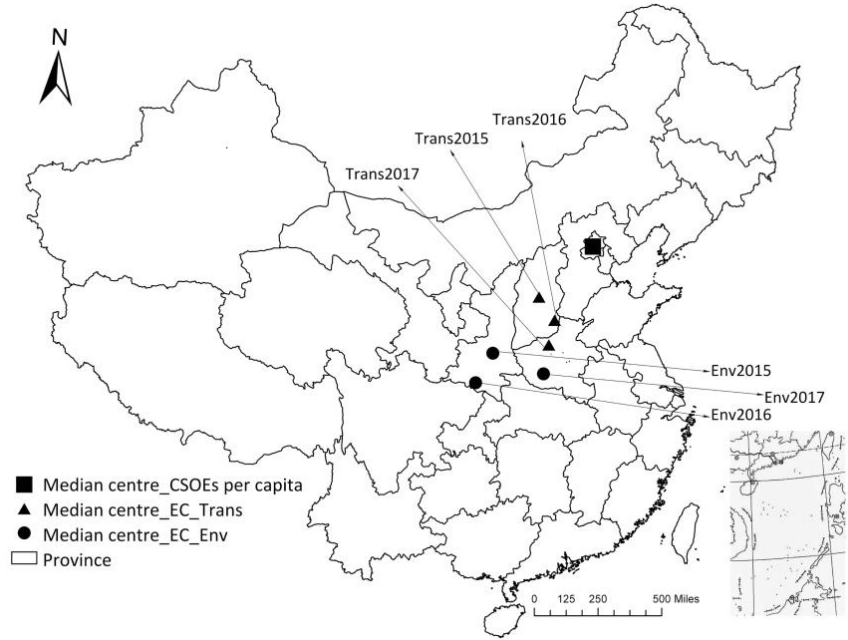

Figure 9(b)

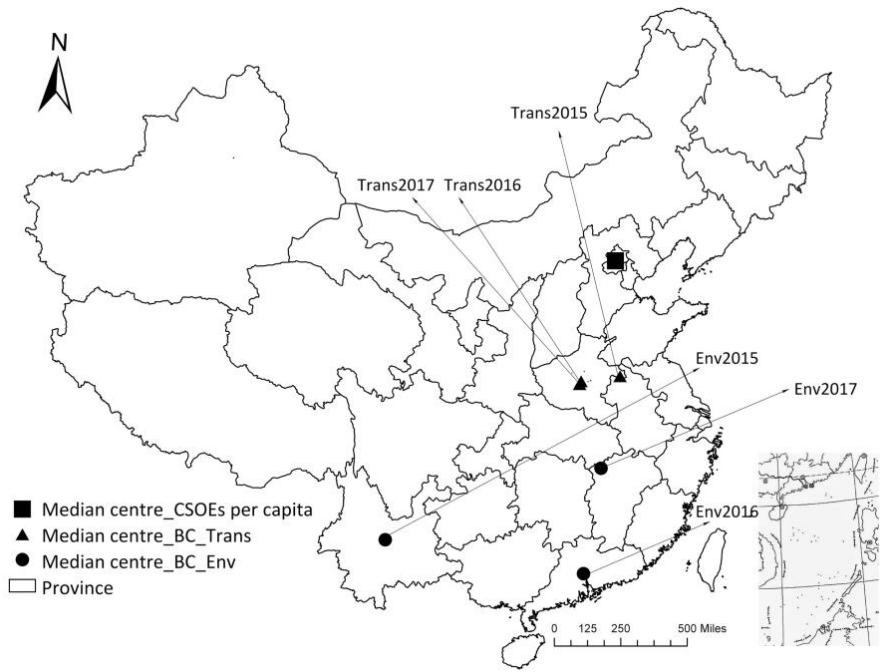

**Figure 9. Central tendency of influence and control power of COSEs in transportation and environmental protection PPP networks vs major CSOEs in China, 2015-2017**
Figure notes: (a) The median centres of influence of CSOEs in transportation and environmental protection PPP networks. (b) The median centres of control power of CSOEs in transportation and environmental protection PPP networks.

Because of the higher reliance and dependence on CSOEs in the transportation sector, as predicted by RDT, CSOEs exert a higher control power imbalance in the transportation sector than in the environmental protection sector.

From a temporal perspective and corresponding to the second hypothesis, CSOEs' influence and control power experienced a rise in 2017 for both transportation and environmental protection sectors. Because of the sudden surge of PPP projects and amounts, China's central governments tightened qualifications and approval requirements of PPP projects since 2017. The tightened qualifications and approval requirements of PPP projects may further strengthen CSOEs' influence and control power in the network in 2017, because they have competitive advantages of the financial capacity, expertise, human capital, credibility, and the like, compared to SOEs and private firms.

From a geographic perspective and corresponding to the third hypothesis, it is suggested that, in the transportation sector, the spatial concertation of CSOEs' influence and control power is more aligned with geographic distribution of CSOEs than in the environmental protection sector. Such alignment may suggest that CSOEs exert greater dominance in the transportation sector and then restrict transportation PPP network in line with provincial CSOEs. In the environmental protection sector, local governments may have greater flexibility and discretion and therefore seek non-CSOEs which have strategic access to key technology and skills.

Research findings presented indicate CSOEs' dominant role in PPP network of both transportation and environmental protection sectors, as a result of CSOEs' superior possession of strategic assets. Policy makers should reduce resource gaps between SOEs and private businesses, and only in so doing, presence and involvement of non-SOEs in China's PPPs can be enhanced. Particularly, the innovative financing tools should be developed for private businesses to obtain adequate financial capital to invest in PPP projects. The profitable projects which are paid by users could be opened to the private partners, while projects with more public interests could be conducted by SOEs.

This paper is important because it develops an analytical framework for understanding and examining the governments' priorities and preferences for collaborating with CSOEs due to their superior access to and control over resources, through a governance network perspective. In addition, SNA is introduced into PPP research, so that each PPP project and collaboration is not isolated or independent from each other. From a network perspective, the interactions among different types of participants in the partnerships are better discerned. This paper also has some limitations and may be expanded in future research. First, it only discusses the PPP networks in the transportation and environmental protection sectors. Abundant cautious may be needed for generalizing the results in this research because it focuses primarily on transportation and environmental protection sectors. Future research directions may focus on comparison of additional sectors and categorizing, in addition, local governments according to their distinct characteristics.

**Notes**

1. The use of both population-scale network data and social network analysis in this paper may contribute to the emerging field 'computational socioeconomics', see the review paper for details on computational socioeconomics (Gao, Zhang, & Zhou, 2019).

2. As the total number of PPP projects before 2015 is almost negligible, here we only explore the trend over the period of 2015-2017.

3. There are 21 CSOEs participating in PPP projects in the transportation and environmental protection sectors. Subsidiaries of CSOEs are included as separate enterprises. In total, there are 175 CSOEs and their subsidiaries acting as societal capital organizations in the two sectors. Eight major CSOEs are selected as more than three of their subsidiaries participate in PPP projects. The proportion of subsidiaries participating in PPPs from those eight CSOEs is near 85%.

# Appendix – Mean comparison statistics for Figures 7-8

Table A.1 – for Figure 7

**Table A.1 The mean-comparison test results of eigenvector centrality of different types of actors in PPP networks, 2015-2017**

| Year | Group1 | Group2 | Transportation | | Environmental protection | |
|---|---|---|---|---|---|---|
| | | | p | p.signif | p | p.signif |
| 2017 | Govt | CSOEs | 6.24E-30 | **** | 6.06E-07 | **** |
| 2017 | Govt | Private | 6.53E-01 | ns | 1.80E-01 | ns |
| 2017 | Govt | SOEs | 4.19E-01 | ns | 9.67E-01 | ns |
| 2017 | CSOEs | Private | 2.68E-24 | **** | 1.08E-05 | **** |
| 2017 | CSOEs | SOEs | 1.47E-21 | **** | 1.12E-05 | **** |
| 2017 | SOEs | Private | 2.19E-01 | ns | 2.70E-01 | ns |
| 2016 | Govt | CSOEs | 5.06E-18 | **** | 1.31E-02 | * |
| 2016 | Govt | Private | 9.96E-01 | ns | 2.29E-01 | ns |
| 2016 | Govt | SOEs | 1.79E-01 | ns | 5.97E-02 | ns |
| 2016 | CSOEs | Private | 5.06E-15 | **** | 7.82E-02 | ns |
| 2016 | CSOEs | SOEs | 7.47E-11 | **** | 6.48E-01 | ns |
| 2016 | SOEs | Private | 1.96E-01 | ns | 2.37E-01 | ns |
| 2015 | Govt | CSOEs | 3.92E-10 | **** | 8.24E-02 | ns |
| 2015 | Govt | Private | 4.46E-01 | ns | 1.15E-01 | ns |
| 2015 | Govt | SOEs | 5.87E-02 | ns | 1.47E-02 | * |
| 2015 | CSOEs | Private | 3.21E-07 | **** | 6.16E-01 | ns |
| 2015 | CSOEs | SOEs | 1.60E-03 | ** | 7.95E-01 | ns |
| 2015 | SOEs | Private | 3.80E-01 | ns | 4.25E-01 | ns |

Note: ns, Non-significant, *$p < 0.05$, **$p < 0.01$, ***$p < 0.001$, ****$p < 0.0001$.

Table A.2 – for Figure 8

**Table A.2 The mean-comparison test results of betweenness centrality of different types of actors in PPP networks, 2015-2017**

| Year | Group1 | Group2 | Transportation | | Environmental protection | |
|---|---|---|---|---|---|---|
| | | | p | p.signif | p | p.signif |
| 2017 | Govt | CSOEs | 6.63E-16 | **** | 9.45E-05 | **** |
| 2017 | Govt | Private | 7.21E-01 | ns | 8.20E-01 | ns |
| 2017 | Govt | SOEs | 1.58E-02 | * | 1.25E-01 | ns |
| 2017 | CSOEs | Private | 6.95E-12 | **** | 9.20E-05 | **** |
| 2017 | CSOEs | SOEs | 1.68E-07 | **** | 6.15E-02 | ns |
| 2017 | SOEs | Private | 1.94E-02 | * | 1.06E-01 | ns |
| 2016 | Govt | CSOEs | 1.92E-08 | **** | 1.01E-01 | ns |
| 2016 | Govt | Private | 2.56E-01 | ns | 9.94E-01 | ns |
| 2016 | Govt | SOEs | 1.24E-01 | ns | 1.05E-01 | ns |
| 2016 | CSOEs | Private | 2.02E-08 | **** | 1.13E-01 | ns |
| 2016 | CSOEs | SOEs | 1.63E-04 | *** | 9.70E-01 | ns |
| 2016 | SOEs | Private | 1.81E-02 | * | 1.17E-01 | ns |
| 2015 | Govt | CSOEs | 7.93E-04 | *** | 7.82E-01 | ns |
| 2015 | Govt | Private | 5.92E-01 | ns | 7.56E-01 | ns |
| 2015 | Govt | SOEs | 4.84E-01 | ns | 5.22E-01 | ns |
| 2015 | CSOEs | Private | 1.22E-03 | ** | 6.54E-01 | ns |
| 2015 | CSOEs | SOEs | 2.37E-02 | * | 4.70E-01 | ns |
| 2015 | SOEs | Private | 2.76E-01 | ns | 7.03E-01 | ns |

Note: ns, Non-significant, *$p < 0.05$, **$p < 0.01$, ***$p < 0.001$, ****$p < 0.0001$.